\documentclass{article}
\usepackage{amsmath}
\usepackage{amssymb}
\usepackage{graphicx}
\usepackage{listings}
\usepackage{enumitem}

\usepackage{algorithm, algpseudocode}
\usepackage{float}
\usepackage{bm}
\usepackage{subfig}
\usepackage{amsmath,amssymb,mathrsfs}
\usepackage{bbm}
\usepackage{url}
\usepackage{wrapfig}
\usepackage{multirow}
\usepackage[normalem]{ulem}
\usepackage{booktabs}
\usepackage{titlesec}
\usepackage[hidelinks]{hyperref}

\usepackage{etoolbox}
\usepackage{url}
\usepackage[most]{tcolorbox}
\usepackage{xcolor}
\newcommand{\qedwhite}{\hfill \ensuremath{\Box}}

\newcommand{\lp}{\left(}
\newcommand{\rp}{\right)}
\newcommand{\lb}{\left[}
\newcommand{\rb}{\right]}
\newcommand{\lbp}{\left\{}
\newcommand{\rbp}{\right\}}
\newcommand{\lba}{\left\lvert}
\newcommand{\rba}{\right\rvert}
\newcommand{\lV}{\left\lVert}
\newcommand{\rV}{\right\rVert}
\newcommand{\mv}{\middle\vert}

\newcommand{\mcal}{\mathcal}

\newcommand{\mbb}{\mathbb}
\newcommand{\msf}{\mathsf}

\newcommand{\ra}{\rightarrow}

\newcommand{\lan}{\langle}
\newcommand{\ran}{\rangle}

\newcommand{\eqDef}{\triangleq}

\newcommand{\E}{\mathbb{E}}
\newcommand{\Var}{\mathsf{Var}}

\newcommand{\argmax}{\mathop{\mathrm{argmax}}}

\usepackage{enumitem }

\usepackage{natbib}   
\newtheorem{theorem}{Theorem}[section]
\newtheorem{proposition}[theorem]{Proposition}
\newtheorem{lemma}[theorem]{Lemma}
\newtheorem{corollary}[theorem]{Corollary}

\newtheorem{assumption}[theorem]{Assumption}

\newtheorem{remark}[theorem]{Remark}

\title{$L_q$ Lower Bounds on Distributed Estimation via Fisher Information}
\author{Wei-Ning Chen \\
\texttt{wnchen@stanford.edu}
\and
Ayfer \"Oz\"gur\\
\texttt{aozgur@stanford.edu}} 
\date{
Stanford University
} 

\begin{document}
\maketitle
\begin{abstract}
    Van Trees inequality, also known as the Bayesian Cram\'er-Rao lower bound, is a powerful tool for establishing lower bounds for minimax estimation through Fisher information. It easily adapts to different statistical models and often yields tight bounds. Recently, its application has been extended to distributed estimation with privacy and communication constraints where it yields order-wise optimal minimax lower bounds for various parametric tasks under squared $L_2$ loss.
   
   However, a widely perceived drawback of the van Trees inequality is that it is limited to squared $L_2$ loss. The goal of this paper is to dispel that perception by introducing a strengthened version of the van Trees inequality that applies to general $L_q$ loss functions by building on the Efroimovich's inequality -- a lesser-known entropic inequality dating back to the 1970s. We then apply the generalized van Trees inequality to lower bound $L_q$ loss in distributed minimax estimation under communication and local differential privacy constraints. This leads to lower bounds for $L_q$ loss that apply to sequentially interactive and blackboard communication protocols. Additionally, we show how the generalized van Trees inequality can be used to obtain \emph{local} and \emph{non-asymptotic} minimax results that capture the hardness of estimating each instance at finite sample sizes.
\end{abstract}

\tableofcontents

\section{Introduction}\label{sec:introduction}
For a real-valued parameter $\theta\in \Theta$ and an observation $X \sim P_\theta$, the basic question of parametric statistics is how well one can estimate $\theta$ from $X$ under a given loss function $\ell( \hat{\theta} - \theta)$. The Fisher information $I_X(\theta)$ plays a crucial role in this context by capturing the \emph{local} hardness of estimating $\theta$ from $X$, with implications both asymptotically \citep{hajek1961some, van2000asymptotic} and non-asymptotically \citep{cramer1999mathematical, rao1992information, van2004detection, gill1995applications, efroimovich1980information, aras2019family}. For example, the famous Cram\'er-Rao lower bound \citep{cramer1999mathematical, rao1992information} states that the squared $L_2$ estimation error (a.k.a. the mean squared error) of any unbiased estimator is lower bounded by the inverse of the Fisher information. The unbiasedness assumption on the estimator can be removed by assuming that the parameter $\theta$ is distributed according to a prior on $\Theta$. This is known as the Bayesian Cram\'er-Rao lower bound or the van Trees Inequality \citep{van2004detection, gill1995applications}. By strategically constructing the prior, this approach can also be used to prove lower bounds for minimax estimation, where the goal is to minimize the worst-case error of the estimator over the parameter space $\Theta$.

The classic parametric estimation tasks have gained renewed popularity over recent years, driven by the prevalence of modern datasets often generated and stored on local devices, such as in federated learning (FL) and analytics (FA) \citep{mcmahan2016communication, mcmahan2017learning, kairouz2019advances}. In these settings, the collection and utilization of decentralized data encounter various resource constraints, including communication and privacy considerations. These constraints have raised the question of how well one can estimate the unknown parameter $\theta\in \Theta$ from a processed observation $Y$, which corresponds to the output of a privatization and/or compression mechanism applied to $X\sim P_\theta$. In \citet{barnes2019fisher, barnes2019lower, barnes2020fisher}, the classical Fisher information framework has been extended to the case of privacy and communication (or compression) constraints. This approach first upper bounds the Fisher information from a differentially private and/or compressed sample $Y$ and then uses the van Trees inequality to lower bound the minimax squared $L_2$ error of distributed estimation.  This, in a unified fashion, leads to order-optimal lower bounds for various parametric tasks under privacy and/or communication constraints, including two tasks that have been of significant interest in the recent literature: distributed mean estimation \citep[see]{duchi2013local, gandikota2021vqsgd, an2016distributed, zhang2012communication, duchi2018right, duchi2019lower, asi2022optimal,agarwal2018cpsgd,chen2020breaking, feldman2021lossless, shah2022optimal, asi2023fast, isik2023exact} and discrete distribution (or frequency) estimation \citep[see]{han2018distributed, han2018geometric,erlingsson14rappor, kairouz16, wang2017locally, ye2017optimal,acharya2019hadamard, acharya2019inference, chen2020breaking, feldman2021lossless, feldman2022private}. 
While order-optimal minimax lower bounds for distributed estimation under information constraints can also be derived by using other techniques, e.g., leveraging strong data-processing inequalities in \citet{duchi2013local, zhang2013information, garg2014communication,  braverman2016communication, duchi2019lower, asoodeh2022contraction} or employing methods based on Le Cam, Fano, or Assaud \citep{acharya2019inference, acharya2019inference2, acharya2020unified}, the constrained Fisher information approach  has several distinct advantages: 
 \begin{itemize}
    \item It is relatively straightforward and applies to various parametric models in a unified fashion, e.g., mean estimation or distribution estimation. In contrast, methods based on Le Cam, Fano, or Assaud are more versatile in that they can be potentially adopted to diverse statistical problems beyond parametric settings, e.g., testing, but the construction of worst-case instances tends to be intricate and is heavily tied to the specific problem structure.
    
    \item The Fisher information approach elucidates a clear relationship between the tail behavior of the score function of the parametric model and error rates under privacy/communication constraints.
    
    \item  Leveraging the chain rule of Fisher information, the Fisher information approach naturally extends to sequentially interactive communication protocols or even fully interactive blackboard protocols \citep{kushilevitz1997communication}.
  
    \item Finally, Fisher information is inherently a local measure of hardness. For example, the influential H\'ajek-Le Cam's local asymptotic minimax (LAM) theorem asserts that, for a general ball-shaped loss function $\ell(\cdot)$, the \emph{asymptotic} estimation error at $\theta\in\Theta$ is at least $\E\left[ \ell(Z)\right]$, where $Z \sim \mathcal{N}\left(0, I^{-1}_X(\theta)\right)$ \citep{hajek1961some, van2000asymptotic}. Using the van Trees inequality with a carefully constructed prior, \citet{chen2021pointwise} shows that Fisher information can be used to prove  \emph{local} and \emph{non-asymptotic} minimax lower bounds that capture the hardness of estimating each instance of the problem at finite sample size. To the best of our knowledge, this is the only example of \emph{non-asymptotic local} lower bounds for distributed estimation, combining the power of H\'ajek-Le Cam type results to capture \emph{local} hardness with the \emph{non-asymptotic} nature of global minimax lower bounds that have been almost exclusively the focus of the existing distributed estimation literature.
 \end{itemize}
 
However, a widely perceived drawback of the Fisher information approach, both for classical parametric statistics as well as its extension to the distributed setting with privacy and communication constraints, is that its applicability is limited to squared $L_2$ error. For example, \citet{tsybakov2004introduction} acknowledges several advantages of van Trees inequality, such as its relative simplicity in application and ability to establish sharp bounds, and also notes \emph{``a limitation is that the van Trees inequality applies only to the squared loss function.''} The goal of this paper is to dispel this perception by showing how Fisher information can be used to prove lower bounds for any $L_q$ loss for $1\leq q<\infty$. We do this by building on a less-known entropic inequality called Efroimovich's inequality \citep{efroimovich1980information} highlighted in recent works \citet{aras2019family,lee2022new}. Our paper makes the following contributions:
\begin{itemize}
    \item We prove a van Trees type inequality for $L_q$ loss by combining Efroimovich's inequality with a maximum entropy argument.\footnote{In \citet{aras2019family,lee2022new}, the authors comment that Efroimovich's inequality can be potentially used to bound loss functions beyond $L_2$. However, we have not been able to find an explicit result in this direction.}
    
    \item We leverage this generalized van Trees inequality to establish global minimax lower bounds for various distributed estimation tasks under $L_q$ loss. As an immediate consequence, this yields lower bounds for $L_q$ loss applicable to sequentially interactive and blackboard communication protocols. Our approach not only recovers previous lower bounds presented in \citet{acharya2020unified} for sequentially interactive models in a cleaner and more straightforward manner, but also extends to fully interactive blackboard communication protocols.
    
    \item We show how the generalized van Trees inequality can be used to derive  \emph{local non-asymptotic} minimax lower bounds for distribution estimation under $L_q$ loss extending the approach of \citet{chen2021pointwise}. This \ emph {local} lower bounds match the performance of a scheme previously developed in \citet{chen2021pointwise} and establish its instance-optimality under $L_1$ loss. The results extend to $L_q$ loss. 
\end{itemize}

\paragraph*{Organization.} The rest of the paper is organized as follows. We recap van Trees inequality and extend it to $L_q$ loss via Efroimovich inequality in Section~\ref{sec:preliminaries}. In Section~\ref{sec:global_minimax}, we leverage the generalized van Trees inequality to establish global minimax lower bounds for various distributed estimation tasks under $L_q$ loss. In Section~\ref{sec:local_minimax}, we extend our focus to derive \emph{local} minimax lower bounds for distribution estimation tasks and provide schemes that achieve the lower bounds pointwisely. Finally, we summarize the work in Section~\ref{sec:conclusion}.


\section{Preliminaries}\label{sec:preliminaries}
In this section, we recap van Trees inequality and Efroimovich inequality and refer interested readers to \cite{aras2019family} and \cite{lee2022new} for more details and recent advances. Throughout the paper, we consider the following parametric model: let
$ \lbp P_\theta \mv \theta \in \Theta\rbp$ with $\Theta \subseteq \mbb{R}^d$ be a  family of probability measures over $\mcal{X}$ with a dominating $\sigma$-finite measure $\lambda$ such that the density:
$$ dP_\theta(\cdot) = f(\cdot; \theta) d\lambda(\cdot) $$
exists. We make the following assumption on the model:
\begin{assumption}\label{assumption:regularity}
    The density function $f(x; \theta)$ is differentiable for $x \in \mcal{X}$ $\lambda$-a.e. In addition, $f$ satisfies
    $$ \int_{\mcal{X}} \nabla_\theta f(x;\theta) d\lambda(x) = 0, $$
    for all $\theta \in \Theta$.
\end{assumption}
Assumption~\ref{assumption:regularity} is a common regularity condition in the Cram\'er-Rao type bounds, allowing for exchanging the differentiation and integration. For a prior distribution $\pi$ over $\Theta$, the information theorists' Fisher information is defined as
\begin{align}\label{def:fisher_info_j}
    J(\pi) \eqDef \int_{\mbb{R}^d} \frac{\lba \nabla_\theta \pi(\theta) \rba^2}{\pi(\theta)},
\end{align}
if it exists. On the other hand, the Fisher information matrix of $\lbp P_\theta |\theta \in \Theta\rbp$ is defined as
\begin{align}
    [I_X(\theta)]_{ij} \eqDef \int_{\mcal{X}} \frac{\frac{\partial}{\partial \theta_i}f(x; \theta) \cdot \frac{\partial}{\partial \theta_j}f(x; \theta)}{f(x; \theta)} d\lambda(x).
\end{align}

Van Trees inequality states that the $L_2$ estimation error for estimating $\theta \in \Theta$, given a prior distribution $\pi$ over $\Theta$, is lower bound by the inverse Fisher information:
\begin{theorem}[\cite{van2004detection, gill1995applications}]\label{thm:van_trees}
    Let $X \sim P_\theta$ and $\theta \sim \pi$ for some prior distribution $\pi$. Let Assumption~\ref{assumption:regularity} hold. Then, for any prior distribution $\pi$ on $\Theta$ for which the information theorist's Fisher information $J(\pi)$
    exists, it holds that
    \begin{equation}
        \E\lb \lV \theta - \hat{\theta} \rV^2_2 \rb \geq \frac{d}{\msf{det}\lp I_X(\theta) + J(\pi) \rp^{1/d}}.
    \end{equation}
\end{theorem}

While van Trees inequality only applies to the $L_2$ error, the following Efroimovich inequality can be used to establish lower bounds for more general loss functions.
\begin{theorem}[\cite{efroimovich1980information}]\label{thm:efroimovich}
    Under the assumptions of Theorem~\ref{thm:van_trees}, we also have
    \begin{equation}
        \frac{1}{2\pi e}e^{\frac{2}{d}h(\theta | X)} \geq \frac{1}{\msf{det}\lp I_X(\theta) + J(\pi) \rp^{1/d}},
    \end{equation}
    where $h(\theta | X)$ is the conditional (differential) entropy.
\end{theorem}

To see how Theorem~\ref{thm:efroimovich} implies the classical van Trees inequality, observe that
\begin{align}\label{eq:entropy_upper_bound}
    &\frac{2}{d}h(\theta | X) \leq \frac{2}{d}h(\theta | \hat{\theta}(X)) \leq \frac{2}{d}h(\theta - \hat{\theta}) \\
    &\leq \log\lp 2\pi e \cdot \msf{det}\lp \E\lb (\hat{\theta}-\theta)(\hat{\theta}-\theta)^\intercal \rb\rp^{1/d} \rp \nonumber\\
    & \leq \log\lp 2\pi e \cdot \frac{1}{d}\E\lb \lV \hat{\theta}-\theta\rV^2_2 \rb\rp. \nonumber
\end{align}
The last two inequalities follow from the fact that the Gaussian distribution maximizes entropy for fixed second moments and the AM-GM inequality. Rearranging yields Theorem~\ref{thm:van_trees}.

Through Efroimovich’s inequality, one can easily extend van Trees inequality to $L_1$ loss: following \eqref{eq:entropy_upper_bound}, we obtain
\begin{align*}
    \frac{2}{d}h(\hat{\theta} - \theta) &\leq \frac{2}{d}\sum_i h(\hat{\theta}_i - \theta_i) 
    \leq \frac{2}{d}\sum_i \log\lp 2e\E\lb \lba \theta_i - \hat{\theta}_i \rba \rb\rp \\
    &\leq 2\log\lp \frac{2e}{d}\sum_i\E\lb \lba \theta_i - \hat{\theta}_i \rba \rb \rp \\
    &= \log\lp \lp \frac{2e}{d}\E\lb \lV \hat{\theta} - \theta \rV_1 \rb \rp^2 \rp.
\end{align*}
The last two inequalities follow from (1) the fact that the Laplace distribution maximizes entropy for a fixed mean absolute error, and (2) the AM-GM inequality. Applying Theorem~\ref{thm:efroimovich}, we immediately obtain the following lower bound for $L_1$ loss:
    \begin{align}\label{eq:l1_lower_bound}
        \E\lb \lV \hat{\theta}-\theta \rV_1 \rb \geq \sqrt{\frac{\pi}{2e}}\frac{d}{\msf{det}\lp I_X(\theta) + J(\pi) \rp^{1/2d}}.
    \end{align}

The above lower bound is order-wise optimal. To see this, suppose that there exists an order-optimal estimator $\hat{\theta}^*(X)$ such that
$$ \E\lb \lV \theta - \hat{\theta}^* \rV^2_2 \rb = C\cdot \frac{d}{\msf{det}\lp I_X(\theta) + J(\pi) \rp^{1/d}}. $$
Then, the Cauchy–Schwartz inequality yields
\begin{align*}
    \E\lb \lV \theta - \hat{\theta}^* \rV_1 \rb \leq \sqrt{d \cdot \E\lb \lV \theta - \hat{\theta}^* \rV^2_2 \rb} = \sqrt{C}\cdot \frac{d}{\msf{det}\lp I_X(\theta) + J(\pi) \rp^{1/2d}}.
\end{align*}
Since it is well-known that the $L_2$ lower bound can be achieved, this argument implies the $L_1$ lower bound is also tight up to a constant. More generally, we can prove the following van Trees inequality for any $L_q$ loss:
\begin{theorem}[Generalized van Trees Inequality]\label{thm:generalized_van_trees}
Let $q \geq 1$ and let $C_{\msf{ME}}(q)$ be the partition function of the $L_q$ max-entropy distribution, formally defined as\footnote{One can verify that $C_\msf{ME}(2) = \sqrt{2\pi e}$ and $C_\msf{ME}(1) = \sqrt{2e}$.} 
$$  C_{\msf{ME}}(q)\eqDef 2e^{\frac{1}{q}}\Gamma\lp \frac{1}{q}\rp q^{\frac{1}{q}-1}.$$
Then, it holds that
    \begin{align*}
        &\E_{\theta,X}\lb \lV \hat{\theta}(X)-\theta \rV_q^q \rb
        \geq \lp \frac{\sqrt{2\pi e}}{C_{\msf{ME}}(q)}\rp^{q} \frac{d}{\msf{det}\lp I_X(\theta) + J(\pi) \rp^{\frac{q}{2d}}}.
    \end{align*}
\end{theorem}

We can simplify the lower bound with the following AM-GM inequality
$$ \msf{det}(A)^{1/d} \leq \frac{\msf{Tr}(A)}{d} \text{ for any PSD } A \in \mbb{R}^{d\times d}, $$
obtaining the following simpler but slightly weaker form:
\begin{proposition}\label{prop:generalized_van_trees}
    Let $q \geq 1$ and $C_{\msf{ME}}(q)$ be defined as in Theorem~\ref{thm:generalized_van_trees}.
Then, it holds that
    \begin{align*}
        \E_{\theta, X}\lb \lV \hat{\theta}(X)-\theta \rV_q^q \rb \geq \lp \frac{\sqrt{2\pi e}}{C_{\msf{ME}}(q)}\rp^{q} \cdot \frac{d^{1+\frac{q}{2}}}{\msf{Tr}\lp I_X(\theta) + J(\pi) \rp^{\frac{q}{2}}}.
    \end{align*}
\end{proposition}

\section{Distributed Estimation under Information Constraints}\label{sec:global_minimax}
The  generalized van Trees inequality (in the form stated in Proposition~\ref{prop:generalized_van_trees}), can be combined with an upper bound on the trace of the  Fisher information matrix to obtain a lower bound on the minimax $L_q$ loss achievable in a distributed setting under privacy and communication constraints. This approach is rather straightforward and is outlined in \cite{barnes2019lower, barnes2020fisher, barnes2019fisher}, where authors also develop the necessary upper bounds on the trace of the Fisher information matrix under privacy and communication constraints. In this section, we overview how privacy and communication constraints are mathematically modeled in a distributed estimation setting  and state the corresponding lower bounds under $L_q$ loss.

\subsection{Problem Setup}
The general distributed statistical task we consider in this paper can be formulated as follows. Each one of the $n$ clients observes a local sample $X_i \sim P_\theta$, processes it via a local channel (i.e., a randomized mapping), and then sends a message $Y_i \in \mcal{Y}$ to the server, which, upon receiving $Y^n$, aims to estimate the unknown parameter $\theta$.

At client $i$, the message $Y_i$ is generated via a \emph{sequential} communication protocol; that is, samples are communicated sequentially by broadcasting the communication to all nodes in the system, including the server. Therefore, the encoding function $W_i$ of the $i$-th client can depend on all previous messages $Y_1,...,Y_{i-1}\in\mathcal{Y}$. Formally, it can be written as a randomized mapping (possibly using shared randomness across participating clients and the server) of the form $Y_i \sim W_i(\cdot| X_i, Y^{i-1})$. 

\paragraph*{$b$-bit communication constraint} The $b$-bit communication constraint restricts $\lba\mcal{Y}\rba \leq 2^b$, ensuring the local message can be described in $b$ bits.

\begin{remark}
    While we only overview the $b$-bit sequentially interactive communication model, our results extend to the $b$-bit blackboard communication protocol \cite{kushilevitz1997communication},where each node is allowed to write $b$-bits in total on a publicly seen blackboard in a randomized order that can depend on the samples. The blackboard model allows for much more interaction between the nodes as compared to the sequential model (e.g. the protocol can start with one of the nodes writing a single bit on the blackboard and the second node to write a bit can depend on the value of the first written bit etc.). The results in Corollary~\ref{cor:b_bit_global_lower_bounds} trivially apply to these more powerful communication protocols simply because the Fisher information bounds in \cite{barnes2019lower, barnes2020fisher}  are proven under this more general model.
\end{remark}

\paragraph*{$\varepsilon$-local differential constraint} The $\varepsilon$-LDP constraint requires that, for any $i\in [n], \, x_i, x'_i \in \mcal{X}, \, y^{i-1} \in \mcal{Y}^{i-1}, \, \mcal{S} \in \sigma\lp Y_i \rp$, it holds that
\begin{align}
    \frac{W_i(\mcal{S}| x_i, y^{i-1})}{W_i(\mcal{S}| x'_i, y^{i-1})} \leq e^\varepsilon.
\end{align}
It ensures that any adversary who observes the output $Y_i$ and the context $Y^{i-1}$ cannot infer the local sample $X_i$.

As a special case, when $W_i$ depends only on $X_i$ and is independent of the other messages $Y^{i-1}$ for all $i$ (i.e. $W_i(\cdot | X_i, Y^{i-1}) = W_i(\cdot | X_i)$), we say the corresponding protocol is \emph{non-interactive}. 
Finally, we call the tuple $\lp W^n, \hat{\theta}(Y^n)\rp$ an estimation scheme, where $\hat{\theta}\lp Y^n \rp$ is an estimator of $\theta$. We use $\Pi_{\msf{seq}}$ and $\Pi_{\msf{ind}}$ to denote the collections of all sequentially interactive and non-interactive schemes, respectively. The goal is to lower bound the global minimax $L_q$ risk:
	$$ \msf{R}\lp L_q, \Theta \rp \eqDef\inf_{\lp W^n, \hat{\theta} \rp \in \Pi_{\msf{seq}}}\sup_{\theta \in \Theta}\E\lb \left\lVert \theta - \hat{\theta}\lp Y^n \rp \right\rVert^q_q \rb $$
and construct  schemes in $\Pi_{\msf{seq}}$ that match these lower bounds.

\subsection{Global Minimax Lower Bounds}
 Combining Proposition~\ref{prop:generalized_van_trees} with the Fisher information bounds in \cite{barnes2019lower}, we obtain the following lower bounds on the $L_q$ risk of common statistical models, including the discrete distribution estimation and the Gaussian mean estimation.
\begin{corollary}[Estimation under $b$-bit constraint]\label{cor:b_bit_global_lower_bounds}
    Let $\kappa(q) \eqDef  \lp \frac{\sqrt{2\pi e}}{C_\msf{ME}(q)} \rp^q$. Then the following lower bounds hold:
    \begin{itemize}
        \item Gaussian location model: let $X \sim \mcal{N}(\theta, \sigma^2I_d)$ with $[-B, B] \subset \Theta$. For $nB^2\min\lp b, d\rp \geq d\sigma^2$, we have
        \begin{align}\label{eq:gaussian_location_global_lower_bound}
            \msf{R}\lp L_q, \Theta \rp \gtrsim d\kappa(q) \max\lbp \lp \frac{d\sigma^2}{nb}\rp^{\frac{q}{2}} , \lp \frac{\sigma^2}{n} \rp^{\frac{q}{2}}\rbp.
        \end{align}

        \item Gaussian covariance estimation: Suppose that $X \sim \mcal{N}\lp 0, \msf{diag}(\theta_1,...,\theta_d) \rp$ with $[\sigma^2_\msf{min},  \sigma^2_\msf{max}] \subset \Theta$. Then for $n\lp \sigma^2_\msf{max} - \sigma^2_\msf{min}\rp^2 \min\lp b^2, d \rp \geq d \sigma_\msf{min}^4$, we have
        \begin{align}\label{eq:gaussian_covariance_global_lower_bound}
            \msf{R}\lp L_q, \Theta \rp \gtrsim d\kappa(q) \max\lbp \lp \frac{d\sigma_\msf{min}^4}{nb^2}\rp^{\frac{q}{2}} , \lp \frac{\sigma_\msf{min}^4}{n} \rp^{\frac{q}{2}}\rbp.
        \end{align}

        \item Distribution estimation: Suppose that $\mcal{X} = \lbp 1, 2, ..., d\rbp$ and that $f(x|\theta) = \theta_x$. Let $\Theta = \Delta_d$ be the $d$-dim probability simplex. For $n\min\lp 2^b, d \rp \geq d^2$, we have
        \begin{align}\label{eq:distribution_estimation_global_lower_bound}
           \msf{R}\lp L_q, \Theta \rp \gtrsim d\kappa(q) \max\lbp \lp \frac{1}{n2^b}\rp^{\frac{q}{2}} , \lp \frac{1}{nd} \rp^{\frac{q}{2}}\rbp.
        \end{align}

        \item Product Bernoulli model: suppose that $X \sim \prod_{i=1}^d \msf{Bern}(\theta_i)$. If $\Theta = [0, 1]^d$, then for $n \min\lbp b, d \rbp \geq d$, we have
        \begin{align}\label{eq:dense_product_bernoulli_global_lower_bound}
            \msf{R}\lp L_q, \Theta \rp \gtrsim d\kappa(q) \max\lbp \lp \frac{d}{nb}\rp^{\frac{q}{2}} , \lp \frac{1}{n} \rp^{\frac{q}{2}}\rbp.
        \end{align}
        for some universal constant $C$. On the other hand, $\Theta = \Delta_d$, then for $n \min\lbp 2^b, d \rbp \geq d^2$, we get instead
        \begin{align}\label{eq:sparse_product_bernoulli_global_lower_bound}
            \msf{R}\lp L_q, \Theta \rp \gtrsim d\kappa(q)  \max\lbp \lp \frac{1}{n2^b}\rp^{\frac{q}{2}} , \lp \frac{1}{nd} \rp^{\frac{q}{2}}\rbp.
        \end{align}
    \end{itemize}
\end{corollary}

Similarly, we can prove lower bounds under the $\varepsilon$-local DP model by using the Fisher information upper bounds in \cite{barnes2020fisher}. We defer this to Corollary~\ref{cor:ldp_global_lower_bounds} in Appendix~\ref{sec:ldp_global_lower_bounds}. Corollary~\ref{cor:b_bit_global_lower_bounds} and Corollary~\ref{cor:ldp_global_lower_bounds} recover several existing lower bounds and extend them into the broader blackboard communication models. Specifically, \eqref{eq:gaussian_location_global_lower_bound} and \eqref{eq:gaussian_location_global_lower_bound_ldp} recover the non-sparse setting of in \cite[Theorem~4]{acharya2020unified} for $q < \infty$; \eqref{eq:distribution_estimation_global_lower_bound} and \eqref{eq:dense_product_bernoulli_global_lower_bound_ldp} recover \cite[Corollary~3]{acharya2020unified}; \eqref{eq:dense_product_bernoulli_global_lower_bound}, \eqref{eq:sparse_product_bernoulli_global_lower_bound}, \eqref{eq:dense_product_bernoulli_global_lower_bound_ldp}, and \eqref{eq:sparse_product_bernoulli_global_lower_bound_ldp} recover \cite[Theorem~3]{acharya2020unified} for $q < \infty$.

\subsection{Achievability}
We note that nearly all of the aforementioned lower bounds are order-wise tight\footnote{The only exception is the sub-exponential case \eqref{eq:gaussian_covariance_global_lower_bound}, in which the tightness of the lower bound (even under $L_2$ loss) remains open.}, meaning that they accurately characterize the correct dependence on parameters such as $d$, $n$, $b$, and $\varepsilon$.
\begin{lemma}\label{lemma:global_minimax_upper_bound}
    Assume the same conditions in Corollary~\ref{cor:b_bit_global_lower_bounds}. Then \eqref{eq:gaussian_location_global_lower_bound}, \eqref{eq:distribution_estimation_global_lower_bound}, \eqref{eq:dense_product_bernoulli_global_lower_bound}, \eqref{eq:sparse_product_bernoulli_global_lower_bound} from Corollary~\ref{cor:b_bit_global_lower_bounds} are tight. The same results hold for Corollary~\ref{cor:ldp_global_lower_bounds} in Appendix~\ref{sec:ldp_global_lower_bounds}.
\end{lemma}

For $1\leq q \leq 2$, the upper bounds readily follow from H\"older's inequality. For $q > 2$,  the order-optimal estimation schemes can be constructed via a sample splitting trick in a fashion similar to \cite{han2018distributed, barnes2020fisher}, which yields optimal error performance. Notably, these upper bounds can be achieved through independent protocols, indicating that the global minimax error is not improved by interaction among clients.

\section{Local Minimax Bounds on Discrete Distribution Estimation}\label{sec:local_minimax}
In Section~\ref{sec:global_minimax}, we establish minimax bounds for various distributed statistical estimation models. However, these \emph{global} minimax lower bounds tend to be too conservative and may not accurately reflect the difficulty of estimating each instance $\theta\in\Theta$. This is especially notable in the case of discrete distribution estimation. For example, we would expect the estimation problem to be inherently easier when the underlying distribution we aim to estimate happens to be sparse. Ideally, it is desirable to have estimation schemes that can automatically adapt to the hardness of the underlying instance, i.e. achieve smaller error in easier instances of the problem, without knowing the instance ahead of time. In contrast,  globally minimax optimal estimation schemes are typically designed and tuned for worst-case scenarios and can therefore be too pessimistic, i.e. not able to exploit structures that make the problem instance easier. 

In this section, we focus on the discrete distribution estimation problem and derive \emph{local} bounds with respect to general $L_q$ loss. 
We consider the same setting as in Section~\ref{sec:global_minimax}; however to highlight that the parameter of interest is an instance of the $d$-dim probability simplex, we use $p = (p_1, p_2, ..., p_d) \in \Delta_d$ to denote the unknown parameter ($\theta\in\Theta$). The goal here is to design a scheme $\lp W^n, \hat{p}\lp Y^n \rp\rp$ to minimize the $L_q$ \emph{local} statistical risk:
	$$ \msf{R}\lp L_q, p, \lp W^n, \hat{p} \rp\rp \eqDef \E\lb \left\lVert p - \hat{p}\lp Y^n \rp \right\rVert^q_q \rb \text{ for all } p\in \Delta_d. $$
We mainly focus on the regime $1\ll d \ll n$ and aim to characterize the \emph{pointwise} statistical convergence rates when $n$ is sufficiently large.

\subsection{$b$-bit communication constraint}
We start with the $b$-bit communication constraint. \cite{chen2021pointwise} provides a two-round scheme that achieves the following estimation error:
\begin{theorem}[$L_2$ local minimax error \cite{chen2021pointwise}]\label{thm:local_upper_bound}
    Let $b \leq \lfloor\log_2 d \rfloor$.
    \begin{enumerate}
        \item There exists a sequentially interactive scheme (with a single round of interaction) $\lp W^n, \check{p} \rp \in \Pi_{\msf{seq}}$, such that for all $p \in \Delta_d$, 
\begin{align}\label{eq:l2_rate}
    \msf{R}\lp L_2, p, \lp W^n, \check{p} \rp\rp 
    &\asymp \frac{\lV p \rV_{\frac{1}{2}}+o_n(1)}{n2^b}.
\end{align}

\item There exists a sequentially interactive  scheme (with a single round of interaction) $\lp \tilde{W}^n, \check{p} \rp \in \Pi_{\msf{seq}}$, such that for all $p \in \Delta_d$, 
\begin{align}\label{eq:l1_rate_b_bit}
    \msf{R}\lp L_1, p, \lp \tilde{W}^n, \check{p} \rp\rp 
    \lesssim \sqrt{\frac{\lV p \rV_{\frac{1}{3}}+o_n(1)}{n2^b}}.
\end{align}
    \end{enumerate}
\end{theorem}

Note that the upper bound on the risk is now a function of the unknown $p$. \cite{chen2021breaking} develops the matching lower bound in \eqref{eq:l2_rate} for $L_2$ loss. The optimality of the $L_1$ rate in \eqref{eq:l1_rate_b_bit} has remained open. In this paper, we close this gap by proving the following matching lower bound for general $L_q$ loss. 
\begin{theorem}[$b$-bit $L_q$ local minimax lower bounds]\label{thm:local_b_bit_lower_bound}
    Let $p \in \Delta'_d \eqDef \lbp p \in \Delta_d \mv \frac{1}{2} < p_1 < \frac{2}{3}\rbp$. Then for any $\delta >0, B \geq \sqrt{\frac{\lV p\rV_{{1}/{2}}}{d2^b}}$, as long as $n = \Omega\lp \frac{d^3\log d}{\lV p\rV_{{1}/{2}}} \rp$, it holds that\footnote{Indeed, the lower bound holds for blackboard interactive schemes \cite{kushilevitz1997communication}, a more general class of interactive schemes than $\Pi_{\msf{seq}}$. See \cite{barnes2019lower} for a discussion of blackboard schemes.}
\begin{align*}
    &\inf_{\lp W^n, \hat{p} \rp \in \Pi_{\msf{seq}}} \sup_{p': \lV p' - p\rV_\infty \leq \frac{B}{\sqrt{n}}} \E_{p'}\lb \lV \hat{p}\lp W^n(X^n) \rp - p' \rV^q_q \rb \\
    &\gtrsim \max\lp \frac{C_\delta\lV p \rV_{\frac{q}{q+2}+\delta}}{(n2^b)^{\frac{q}{2}}}, \frac{{\lV p \rV_{\frac{q}{q+2}}}}{(n2^b)^{\frac{q}{2}}\log d}, \frac{ C_\delta \lV p\rV_{q/2+\delta}^{q/2+\delta}}{n^{\frac{q}{2}}}, \frac{\lV p \rV^{q/2}_{q/2}}{n^{\frac{q}{2}}\log d}\rp,
\end{align*}
where $C_\delta \eqDef \lp\delta/(1+\delta)\rp^2$ is a $\delta$-dependent constant.
\end{theorem}
\paragraph*{Sketch of the proof}
The proof is based on the framework introduced in \cite{barnes2019lower} (see also \cite{chen2021pointwise} for the analysis of $\ell_2$ case), where a global upper bound on the quantized Fisher information is given and used to derive the minimax lower bound on the $\ell_2$ error. We extend their results to the local regime and develop a \emph{local} upper bound on the quantized Fisher information around a neighborhood of $p$.

To obtain a local upper bound, we construct an $h$-dimensional parametric sub-model $\Theta^h_{p}$ that contains $p$ and is a subset of $\mcal{P}_{d}$, where $h \in [d]$ is a tuning parameter and will be determined later. By considering the sub-model $\Theta^h_p$, we can control its Fisher information around $p$ with a function of $h$ and $p$. Optimizing over $h \in [d]$ yield an upper bound that depends on $\lV p \rV_{\frac{q}{q+2}}$. Finally, the local upper bound on the quantized Fisher information is translated to a local minimax lower bound on the $L_q$ error via the generalized van Trees inequality ( Theorem~\ref{thm:generalized_van_trees}). We defer the detailed proof to Appendix~\ref{appendix:proof_of_b_bit_lower_bound}. \qedwhite

\begin{remark}
    Note that Theorem~\ref{thm:local_b_bit_lower_bound} complements Theorem~\ref{thm:local_upper_bound} and is nearly tight (up to a $\log d$ factor) for $q=1$ and $q=2$. We believe the upper bound can be easily adapted to match the lower bound for all $q > 1$.
\end{remark}

\subsection{$\varepsilon$-local differential privacy}
Similarly, under the $\varepsilon$-local DP constraint, we prove the following local lower bound:
\begin{theorem}[$\varepsilon$-LDP $L_q$ local minimax lower bounds]\label{thm:local_ldp_lower_bound}
    Let $p \in \Delta'_d \eqDef \lbp p \in \Delta_d \mv \frac{1}{2} < p_1 < \frac{2}{3}\rbp$. Then for any $\delta >0, B \geq \sqrt{\frac{\lV p\rV_{{1}/{2}}}{d\min\lp e^\varepsilon, \lp e^\varepsilon-1\rp^2\rp}}$, as long as $n = \Omega\lp \frac{d^3\log d}{\lV p\rV_{{1}/{2}}} \rp$, it holds that
\begin{align*}
    &\inf_{\lp W^n, \hat{p} \rp \in \Pi_{\msf{seq}}} \sup_{p': \lV p' - p\rV_\infty \leq \frac{B}{\sqrt{n}}} \E_{p'}\lb \lV \hat{p}\lp W^n(X^n) \rp - p' \rV^q_q \rb \\
    &\gtrsim \max\Bigg( \frac{C_\delta\lV p \rV_{\frac{q}{q+2}+\delta}}{\lp n \min\lp e^\varepsilon, \lp e^\varepsilon-1\rp^2\rp \rp^{\frac{q}{2}}}, \\
    & \quad \frac{C_1{\lV p \rV_{\frac{q}{q+2}}}}{\lp n\min\lp e^\varepsilon, \lp e^\varepsilon-1\rp^2\rp\rp^{\frac{q}{2}}\log d},
     \frac{ C_\delta\cdot \lV p\rV_{q/2+\delta}^{q/2+\delta}}{n^{\frac{q}{2}}}, \frac{C_2\lV p \rV^{q/2}_{q/2}}{n^{\frac{q}{2}}\log d}\Bigg),
\end{align*}
for some $C_\delta, C_1, C_2 > 0$.
\end{theorem}
However, unlike the $b$-bit communication constraint, the tightness of Theorem~\ref{thm:local_ldp_lower_bound} is unknown, and a matching achievability scheme currently remains unsettled.

\section{Conclusion and Future Works}\label{sec:conclusion}
In this work, we first point out that the popular van Trees inequality can be generalized to accommodate general loss functions through Efroimovich’s inequality. Subsequently, we extend the application of the generalized van Trees inequality to $L_q$ loss in the context of distributed estimation under communication and local differential privacy constraints. Notably, combining with previous bounds on constrained Fisher information, our results offer a significantly simplified analysis over existing (global) lower bounds and, more importantly, can be used to derive local minimax results that capture the hardness of instances.



\section*{Acknowledgement}
We thank the anonymous Reviewer 2 for pointing out a critical error in the proof of Theorem~\ref{thm:local_b_bit_lower_bound}.
\newpage
\bibliography{references}

\newpage
\appendix
\section{Global Minimax Lower Bounds under LDP}\label{sec:ldp_global_lower_bounds}
Subject to $\varepsilon$-LDP constraints, the following lower bounds on the risk holds:
\begin{corollary}[Distributed estimation under $\varepsilon$-bit local DP]\label{cor:ldp_global_lower_bounds}
    Making the same assumptions as in Corollary~\ref{cor:b_bit_global_lower_bounds}
    \begin{itemize}
        \item Gaussian location model: let $X \sim \mcal{N}(\theta, \sigma^2I_d)$ with $[-B, B] \subset \Theta$. For $nB^2\min\lp \varepsilon, \varepsilon^2, d\rp \geq d\sigma^2$, we have
        \begin{align}\label{eq:gaussian_location_global_lower_bound_ldp}
           \msf{R}\lp L_q, \Theta \rp \gtrsim d\kappa(q)   \lp \frac{d}{n\min\lp \varepsilon, \varepsilon^2\rp}\rp^{\frac{q}{2}}.
        \end{align}

        \item Distribution estimation: Suppose that $\mcal{X} = \lbp 1, 2, ..., d\rbp$ and that $f(x|\theta) = \theta_x$. Let $\Theta = \Delta_d$ be the $d$-dim probability simplex. For $n\min\lp e^\varepsilon, \lp e^{\varepsilon}-1\rp^2, d \rp \geq d^2$, we have
        \begin{align}\label{eq:distribution_estimation_global_lower_bound_ldp}
           \msf{R}\lp L_q, \Theta \rp \gtrsim d\kappa(q) \lp \frac{1}{n \min\lp e^\varepsilon, \lp e^\varepsilon-1 \rp^2, d\rp}\rp^{\frac{q}{2}}.
        \end{align}

        \item Product Bernoulli model: suppose that $X \sim \prod_{i=1}^d \msf{Bern}(\theta_i)$. If $\Theta = [0, 1]^d$, then for $n \min\lbp b, d \rbp \geq d$, we have
        \begin{align}\label{eq:dense_product_bernoulli_global_lower_bound_ldp}
            \msf{R}\lp L_q, \Theta \rp \gtrsim d\kappa(q) \lp \frac{d}{n\lp \varepsilon, \varepsilon^2, d \rp}\rp^{\frac{q}{2}}.
        \end{align}
        On the other hand, $\Theta = \Delta_d$, then for $n \min\lbp 2^b, d \rbp \geq d^2$, we get instead
        \begin{align}\label{eq:sparse_product_bernoulli_global_lower_bound_ldp}
           \msf{R}\lp L_q, \Theta \rp \gtrsim d\kappa(q) \lp \frac{1}{n\min\lp e^\varepsilon, \lp e^\varepsilon-1 \rp^2, d\rp}\rp^{\frac{q}{2}}.
        \end{align}
    \end{itemize}
\end{corollary}

\section{Missing Proofs}

\subsection{Proof of Theorem~\ref{thm:generalized_van_trees}}
We start with the following maximum-entropy lemma.
\begin{lemma}[Maximum entropy under $p$-th moment constraints.]\label{lemma:max_entropy}
    For any random variable $X \in \mbb{R}$, it holds that
    $$ h(X) \leq \log\lp 2e^{\frac{1}{q}}\Gamma\lp \frac{1}{q}\rp q^{\frac{1}{q}-1}\lp\E\lb \lba X\rba^q \rb\rp^{\frac{1}{q}}\rp \eqDef \log\lp C_{\msf{ME}}(q)\cdot\E\lb \lba X\rba^q \rb^{\frac{1}{q}}\rp. $$
\end{lemma}

Now, we proceed with \eqref{eq:entropy_upper_bound}:
\begin{align*}
    \frac{2}{d}h(\hat{\theta} - \theta) &\leq \frac{2}{d}\sum_i h(\hat{\theta}_i - \theta_i) 
    \leq \frac{2}{d}\sum_i \log\lp C_{\msf{ME}}(q)\lp\E\lb \lba\hat{\theta}_i - \theta_i\rba^q \rb\rp^{\frac{1}{q}}\rp \\
    &\leq 2\log\lp C_{\msf{ME}}(q)\lp \frac{1}{d}\sum_i \E\lb \lba\hat{\theta}_i - \theta_i\rba^q \rb\rp^{\frac{1}{q}}\rp \\
    &= \log\lp\lp C_{\msf{ME}}(q)\lp \frac{1}{d} \E\lb\lV \hat{\theta} - \theta\rV^q_q \rb\rp^{\frac{1}{q}}\rp^2\rp.
\end{align*}
Plugging into Theorem~\ref{thm:efroimovich} and rearranging yield the desired result. \qedwhite

\subsection{Proof of Theorem~\ref{thm:local_b_bit_lower_bound}}\label{appendix:proof_of_b_bit_lower_bound}
Before entering the main proof, we first introduce some notation that will be used throughout this section. Let $(p_{(1)}, p_{(2)},...,p_{(d)})$ be the sorted sequence of $p = (p_1, p_2,...,p_d)$ in the non-increasing order; that is, $p_{(i)} \geq p_{(j)}$ for all $i > j$.  Denote $\pi:[d]\ra[d]$ as the corresponding sorting function\footnote{With a slight abuse of notation, we overload $\pi$ so that $\pi\lp (p_1, ..., p_d) \rp \eqDef (p_{\pi(1)},...,p_{\pi(d)})$}, i.e. $p_{(i)} = p_{\pi(i)}$ for all $i \in [d]$. 

\paragraph{Constructing the sub-model $\Theta^h_p$.} 
We construct $\Theta^h_{p}$ by freezing the $(d-h)$ smallest coordinates of $p$ and only letting the largest $(h-1)$ coordinates to be free parameters. Mathematically, let
\begin{equation}
    \Theta^h_{p} \eqDef \lbp \lp \theta_2, \theta_3, ..., \theta_h \rp \mv \pi^{-1}\lp \theta_1, \theta_2,...,\theta_h, p_{(h+1)}, p_{(h+2)},...,p_{(d)} \rp \in \mcal{P}_d\rbp,
\end{equation}
where $\theta_1  = 1 - \sum_{i=2}^h \theta_i - \sum_{i=h+1}^d p_i$ is fixed when $(\theta_2,...,\theta_h)$ are determined. For instance, if $p = \lp \frac{1}{16}, \frac{1}{8}, \frac{1}{2}, \frac{1}{16}, \frac{1}{4}\rp$ (so $d=5$) and $h=3$, then the corresponding sub-model is $$ \Theta^h_p \eqDef \lbp (\theta_2, \theta_3) \mv \lp \frac{1}{16}, \theta_3, \theta_1, \frac{1}{16}, \theta_2 \rp\in \mcal{P}_d \rbp.$$

\paragraph{Bounding the quantized Fisher information.} 
Now recall that under this model, the score function 
$$ S_{\theta}(x) \eqDef \lp S_{\theta_2}(x),...,S_{\theta_h}(x)\rp \eqDef \lp \frac{\partial \log p(x|\theta)}{\partial \theta_2},...,\frac{\partial \log p(x|\theta)}{\partial \theta_h} \rp$$ can be computed as
$$    
S_{\theta_i}(x) = 
\begin{cases}
    \frac{1}{\theta_i}, &\text{ if } x = \pi(i), \, 2 \leq i \leq h\\
    -\frac{1}{\theta_1}, &\text{ if } x = \pi(1)\\
    0, &\text{ otherwise}
\end{cases}
$$

The next lemma shows that to bound the quantized Fisher information, it suffices to control the variance of the score function.
\begin{lemma}[Theorem~1 in \cite{barnes2019lower}]\label{lemma:FI_bdd}
Let $W$ be any $b$-bit quantization scheme and $I_W(\theta)$ is the Fisher information of $Y$ at $\theta$ where $Y \sim  W(\cdot|X)$ and $X\sim p_\theta$. Then for any $\theta \in \Theta \subseteq \mbb{R}^h$,
$$ \msf{Tr}\lp I_W(\theta) \rp \leq \min \lp \msf{Tr}\lp I_X(\theta)\rp, 2^b \max_{\lV u\rV_2\leq 1}\Var\lp \lan u, S_\theta(X)\ran \rp\rp. $$
\end{lemma}
Therefore, for any unit vector $u = (u_2,...,u_h)$ with $\lV u \rV_2=1$, we control the variance as follows:
\begin{align}\label{eq:score_bdd}
    \Var\lp \lan u, S_\theta(X)\ran \rp 
    & \overset{\text{(a)}}{=} \sum_{i=1}^h\theta_i\lp \sum_{j=2}^h u_j S_{\theta_j}(\pi(i)) \rp^2 \nonumber\\
    & = \theta_1\lp \sum_{j=2}^h u_j\rp^2 \lp \frac{1}{\theta_1} \rp^2 + \sum_{i=2}^h \theta_i u_i^2\frac{1}{\theta_i^2}\nonumber\\
    & = \frac{\lp\sum_{j=2}^h u_j\rp^2}{\theta_1} + \sum\limits_{j=2}^h\frac{u_i^2}{\theta_i}\nonumber\\
    & \leq \frac{h}{\theta_1} + \frac{1}{\min_{j \in \lbp 2,...,h\rbp} \theta_j},
\end{align}
where (a) holds since the score function has zero mean. 
This allows us to upper bound $I_W(\theta)$ in a neighborhood around $\theta(p)$, where $\theta(p)$ is the location of $p$ in the sub-model $\Theta^h_p$, i.e. 
$$\theta(p) = (\theta_2(p),...,\theta_h(p))\eqDef (p_{(2)},...,p_{(h)}). $$

In particular, for any $ 0 < B \leq \frac{p_{(h)}}{3}$ and $p \in \mcal{P}'_d \eqDef \lbp p \in \mcal{P}_d \mv \frac{1}{2} < p_1 < \frac{5}{6}  \rbp$, the neighborhood $\mcal{N}_{B, h}(p) \eqDef \theta(p)+[-B, B]^h$ must be contained in $\Theta^h_p$. Next, we control the quantized Fisher information over $\mcal{N}_{B, h}(p)$ as follows.
\begin{enumerate}
    \item For any $\theta' \in \mcal{N}_{B, h}(p)$, it holds that
$$ \theta_1' \geq \theta_1(p) - \frac{hp_{(h)}}{3} \geq \frac{1}{6},$$
where the second inequality holds since 1) $\theta_1(p) = p_{(1)} \geq \frac{1}{2}$ by our definition of $\mcal{P}'_d$, and 2) $\frac{hp_{(h)}}{3} \leq \frac{\sum_{i=1}^h p_{(1)}}{3} \leq \frac{1}{3}$. We also have 
$$ \min_{j \in \{2,...,h\}}\theta_j' \geq \min_{j \in \lbp 2,...,h\rbp}\theta_j(p) - \frac{p_{(h)}}{3} \geq \frac{2p_{(h)}}{3}.$$
Therefore \eqref{eq:score_bdd} implies for any $\theta'\in\mcal{N}_{B,h}(p)$, 
\begin{equation}\label{eq:score_bdd1}
    \Var\lp \lan u, S_\theta'(X)\ran \rp  \leq 6h + \frac{3}{2p_{(h)}}.
\end{equation}

\item By the definition of $S_{\theta_i}(x)$ and Fisher information matrix,
$$ \msf{Tr}\lp I_X(\theta) \rp = \sum_{x}\sum_{i\in [h]} S_{\theta_i}(x)S_{\theta_i}(x)p_\theta(x) = \sum_{i=1}^h \frac{1}{p_{(i)}}. $$
Maximizing $\msf{Tr}\lp I_X(\theta) \rp$ yields 
\begin{equation}\label{eq:score_bdd2}
    \max_{\theta' \in \mcal{N}_{B, h}(p)} \msf{Tr}\lp I_X(\theta) \rp = \sum_{i=1}^h \frac{1}{p_{(i)} - \frac{1}{3}p_{(h)}} \leq \frac{3}{2}\sum_{i}\frac{1}{p_{(i)}} \leq \frac{3}{2}\frac{h}{p_{(h)}}.
\end{equation}
\end{enumerate}

Combining \eqref{eq:score_bdd1} and \eqref{eq:score_bdd2} with Lemma~\ref{lemma:FI_bdd}, we arrive at
$$ \forall \theta' \in \mcal{N}_{B, h}(p), \,\, \msf{Tr}\lp I_W(\theta') \rp \leq \min\lp 2^b\lp 6h+\frac{3}{2p_{(h)}} \rp, \frac{3h}{2p_{(h)}}\rp \leq 2^b\lp 6h+\frac{3}{2p_{(h)}} \rp+ \frac{3h}{2p_{(h)}}. $$

\paragraph{Bounding the $L_q$ error.}
We choose the prior $\bm{\mu}$ on $\mcal{N}_{B, h}(p)$ to minimize the Fisher information, i.e., $\bm{\mu} = \mu^{\otimes h}$ where $\mu$ is the cosine prior introduced in \citet{borovkov, Tsybakov2008}. It is straightforward to see that $\msf{Tr}\lp I(\pi) = \frac{h\pi^2}{B^2} \rp$. Then, applying Proposition~\ref{prop:generalized_van_trees} on $\mcal{N}_{B, h}(p)$ yields
\begin{align}\label{eq:l2_lbd1}
    \sup_{\theta' \in \mcal{N}_{B,h}(p)} \E\lb \lV \hat{\theta} - \theta' \rV^q_q \rb 
    & \geq \frac{h^{1+\frac{q}{2}}}{\lp n2^b\lp 6h+\frac{3}{2p_{(h)}}\rp+\frac{h\pi^2}{B^2} + \frac{3nh}{2p_{(h)}}\rp^{q/2}} \nonumber\\
    & \geq \frac{h^{1+\frac{q}{2}}p_{(h)}^{\frac{q}{2}}}{\lp n2^b\lp 6hp_{(h)}+\frac{3}{2}\rp+\frac{h\pi^2p_{(h)}}{B^2} + \frac{3nh}{2}\rp^{q/2}} \nonumber\\
    & \overset{\text{(a)}}{\geq} \frac{h^{1+\frac{q}{2}}p_{(h)}^{\frac{q}{2}}}{\lp 10n2^b+\frac{10hp_{(h)}}{B^2} + 4nh\rp^{q/2}}.
\end{align}
where the (a) is due to $hp_{(h)} \leq 1$. Now, if we pick $B \geq \sqrt{\frac{hp_{(h)}}{{n2^b}}}$, then \eqref{eq:l2_lbd1} becomes
\begin{equation}\label{eq:l2_lbd2}
    \eqref{eq:l2_lbd1} \geq \frac{h^{1+\frac{q}{2}}p_{(h)}^{\frac{q}{2}}}{\lp 20n2^b + 4nh\rp^{q/2}} \geq \min\lp \frac{h^{1+\frac{q}{2}}p_{(h)}^{\frac{q}{2}}}{\lp 40n2^b\rp^{q/2}} , \frac{h^{1+\frac{q}{2}}p_{(h)}^{\frac{q}{2}}}{\lp 8nh\rp^{q/2}}  \rp = \min\lp \frac{h^{1+\frac{q}{2}}p_{(h)}^{\frac{q}{2}}}{\lp 40n2^b\rp^{q/2}} , \frac{hp_{(h)}^{\frac{q}{2}}}{\lp 8n\rp^{q/2}}  \rp.
\end{equation}

Notice that in order to satisfy the condition $\mcal{N}_{B, h}(p) \subseteq \Theta^h_p$, $B$ must be at most $\frac{p_{(h)}}{3}$, so we have an implicit sample size requirement: $n$ must be at least $\frac{3h}{2^bp_{(h)}}$.

\paragraph{Optimizing over $h$.} 
Finally, we maximize \eqref{eq:l2_lbd2} over $h \in [d]$ to obtain the best lower bound, via the following simple but crucial lemma.
\begin{lemma}\label{lemma:hhp_bdd}
For any $p\in\mcal{P}_d$ and $\delta > 0$, it holds that
\begin{itemize}
    \item $$ \max_{h \in [d]} h^{1+\frac{q}{2}}\cdot p^{\frac{q}{2}}_{(h)} \geq \max\lp C_\delta\lV p \rV_{\frac{q}{q+2}+\delta}, \frac{C{\lV p \rV_{\frac{q}{q+2}}}}{\log d}\rp,$$ for $C_\delta \eqDef \lp\frac{\delta}{1+\delta}\rp^{\frac{2}{1+\delta}} $ and a universal constant $C$ small enough. 
    \item $$ \max_{h\in[d]} \, h \cdot p_{(h)}^{q/2} \geq \max\lp C_\delta\cdot \lV p\rV_{q/2+\delta}^{q/2+\delta},  \frac{C\lV p \rV^{q/2}_{q/2}}{\log d}\rp, $$
    for $C_\delta \eqDef \lp \frac{\delta}{1+\delta} \rp^{\frac{2}{1+\delta}}$ and a universal constant $C$ small enough.
\end{itemize}
\end{lemma}

Picking $h^* = \argmax_{h\in[d]} h^2 p_{(h)}$ and by Lemma~\ref{lemma:hhp_bdd} and $\eqref{eq:l2_lbd2}$, we obtain that for all $\hat{\theta}$
\begin{equation*}
    \sup_{\theta' \in \mcal{N}_{B_n,h^*}(p)} \E\lb \lV \hat{\theta} - \theta' \rV^q_q \rb \geq \max\lp \frac{C_\delta\lV p \rV_{\frac{q}{q+2}+\delta}}{(n2^b)^{\frac{q}{2}}}, \frac{C_1{\lV p \rV_{\frac{q}{q+2}}}}{(n2^b)^{\frac{q}{2}}\log d}, \frac{ C_\delta\cdot \lV p\rV_{q/2+\delta}^{q/2+\delta}}{n^{\frac{q}{2}}}, \frac{C_2\lV p \rV^{q/2}_{q/2}}{n^{\frac{q}{2}}\log d}\rp,
\end{equation*}
as long as $p \in \mcal{P}'_d$ and $ B_n = \sqrt{\frac{h^*p_{(h^*)}}{n2^b}} \overset{\text{(a)}}{\leq} \sqrt{\frac{d}{n2^b\lV p \rV_{\frac{1}{2}}}}$, where (a) holds due to the second result of Lemma~\ref{lemma:hhp_bdd} and $h^*\leq d$. In addition, the sample size constraint that $n$ must be larger than
$\frac{3h^*}{2^bp_{(h^*)}}$ can be satisfied if $n = \Omega\lp \frac{d^3\log d}{2^b \lV p \rV_{\frac{1}{2}}} \rp$ since
    $\frac{h^*}{p_{(h^*)}} \leq \frac{\lp h^*\rp^3\log d}{C \lV p \rV_{\frac{1}{2}}} \leq \frac{d^3\log d}{C \lV p \rV_{\frac{1}{2}}}$,
where the first inequality is due to Lemma~\ref{lemma:hhp_bdd} and the second one is due to $h^* \leq d$. The proof is complete by observing that
$$ \inf_{(W^n, \hat{p})}\sup_{p' : \lV p' -p \rV_\infty\leq B_n} \E\lb \lV \hat{p} - p'\rV^q_q \rb \geq \inf_{\lp W^n, \hat{\theta}\rp}\sup_{\theta' \in \mcal{N}_{B_n,h^*}(p)} \E\lb \lV \hat{\theta} - \theta' \rV^q_q\rb. $$
\qedwhite

\subsection{Proof of Corollary~\ref{cor:b_bit_global_lower_bounds} and Corollary~\ref{cor:ldp_global_lower_bounds}}
The corollaries follow from a direct application of Theorem~\ref{thm:generalized_van_trees}, along with \cite[Corollary~1-4]{barnes2019lower} and \cite[Corollary~1-3]{barnes2020fisher}.

\subsection{Proof of Theorem~\ref{thm:local_ldp_lower_bound}}
The proof follows exactly the same as the Section~\ref{thm:local_b_bit_lower_bound}, except for replacing Lemma~\ref{lemma:FI_bdd} with the following:
\begin{lemma}[Proposition~2 and 3 and in \cite{barnes2020fisher}]\label{lemma:ldp_FI_bdd}
Let $W$ be any $\varepsilon$-local DP scheme and $I_W(\theta)$ is the Fisher information of $Y$ at $\theta$ where $Y \sim  W(\cdot|X)$ and $X\sim p_\theta$. Then for any $\theta \in \Theta \subseteq \mbb{R}^h$,
$$ \msf{Tr}\lp I_W(\theta) \rp \leq \min \lbp \msf{Tr}\lp I_X(\theta)\rp, e^\varepsilon \max_{\lV u\rV_2\leq 1}\Var\lp \lan u, S_\theta(X)\ran \rp, \lp e^\varepsilon - 1\rp^2 \max_{\lV u\rV_2\leq 1}\Var\lp \lan u, S_\theta(X)\ran \rp\rbp. $$
\end{lemma}

\section{Proof of Lemmas}

\subsection{Proof of Lemma~\ref{lemma:max_entropy}}
Let $X$ satisfy the $q$-th moment constraint $\E\lb |X|^q\rb \leq \Delta$. Then, by the maximum entropy principle, the density of $X$ that maximizes the entropy must take the following form:
$$ f(x; \beta) = \frac{\exp\lp - |x|^q/\beta\rp}{\lambda(q, \beta)}, $$
where 
$$ \lambda(q, \beta) \eqDef \int_{\mbb{R}}\exp\lp - |x|^q/\beta\rp dx = \frac{2\beta^{1/q}}{q}\Gamma\lp \frac{1}{q} \rp. $$
Simple calculations yield
\begin{align*}
    \E_{X\sim f(x; \beta)}\lb |X|^q \rb
    = \frac{2}{\lambda(q, \beta)} \int_{0}^\infty x^q \exp\lp -x^q/\beta \rp dx 
    = \frac{2\frac{\beta^{1/q+1}}{q}\Gamma\lp 1+\frac{1}{q} \rp}{\lambda(q, \beta)} 
    = \frac{\beta\Gamma\lp1+\frac{1}{q}\rp}{\Gamma\lp \frac{1}{q} \rp} 
    = \frac{\beta}{q} \leq \Delta.
\end{align*}
On the other hand, the differential entropy can be calculated as 
$$ h(X) = \frac{1}{q} - \log\lp \frac{q}{2\beta^{1/q}\gamma\lp \frac{1}{q} \rp} \rp \leq \frac{1}{q}- \log\lp \frac{p^{1-1/q}}{2\Delta^{1/q}\Gamma\lp \frac{1}{q} \rp} \rp, $$
which establishes the lemma. 

\qedwhite

\subsection{Proof of Lemma~\ref{lemma:hhp_bdd}}
The proof follows from the same arguments as in \cite[Lemma~6.2]{chen2021pointwise} and is omitted here. \qedwhite

\end{document}